 \documentclass[final,3p,times]{elsarticle}

\usepackage{amssymb}
\usepackage{amsmath}
\usepackage{algorithmic}
\usepackage[ruled,linesnumbered]{algorithm2e} 
\usepackage{epsfig}
\usepackage{epstopdf}
\usepackage{subfig}
\usepackage{multirow}
\usepackage{booktabs}

\journal{}

\begin{document}

\begin{frontmatter}



\title{AES-SpMM: Balancing Accuracy and Speed by Adaptive Edge Sampling Strategy to Accelerate SpMM in GNNs}

 \author[SWUST]{Yingchen Song}
 \author[SWUST]{Yaobin Wang\corref{corresponding}}
 \ead{wangyaobin@foxmail.com}
 \author[SWUST]{Yi Luo}
 \author[SWUST]{Huan Wu}
 \author[SWUST]{Pingping Tang}
 
 \cortext[corresponding]{Corresponding author.}
 
 \affiliation[SWUST]{organization={School of Computer Science and Technology},
             addressline={Southwest University of Science and Technology},
             city={Mianyang},
             postcode={621010},
             state={Sichuan},
             country={China}}

\begin{abstract}
Coordinating the design of sampling and sparse-dense matrix multiplication (SpMM) is crucial for accelerating graph neural networks (GNNs). However, due to irrational sampling strategies, existing methods face a trade-off between accuracy and speed. Moreover, as computational optimizations progress, data loading has gradually become the primary bottleneck in GNN inference. To address these issues, we propose AES-SpMM, an adaptive edge sampling SpMM kernel. It considers the relationship between the number of non-zero elements in each matrix row and the shared memory width. The edge sampling scheme is adaptively selected according to the different situations of each row. AES-SpMM reduces the graph size through adaptive edge sampling to fit the GPU's shared memory, lowering the computational cost and enhancing data locality, thus balancing the accuracy and speed of GNN inference. Additionally, we introduce a quantization-based AES-SpMM, which applies quantization and dequantization to feature data in GNNs. This approach significantly reduces data loading time while keeping accuracy loss negligible. We evaluated AES-SpMM with common GNN models and datasets. The results show that AES-SpMM outperforms both the cuSPARSE SpMM kernel and GE-SpMM by up to 25.87$\times$ and 23.01$\times$, respectively, with less than 1\% accuracy loss. Compared to ES-SpMM, it reduces accuracy loss by 3.4\% on average , achieving a 1.31$\times$ speedup. Compared to AES-SpMM, quantization-based AES-SpMM has a maximum accuracy loss of 0.3\% and feature data loading time overhead is reduced by 50.91\%-70.51\%.
\end{abstract}



\begin{keyword}
Graph neural networks\sep Sparse-dense matrix multiplication\sep Sampling\sep Quantization\sep GPU


\end{keyword}

\end{frontmatter}



\section{Introduction}
\label{sec1}

In recent years, graph neural networks (GNNs) have gained significant attention in machine learning due to their effective representation of graph structures in various domains, such as recommender systems~\cite{wu2022graph}, social networks~\cite{zhang2022improving}, biology~\cite{wang2023pros}, and molecular analysis~\cite{cai2022fp}. GNN models data using a graph structure of nodes and edges, with each node associated with a feature vector, and the GNN algorithm operates by edge propagation. Unlike traditional neural networks, GNNs involve feature aggregation operations, which can be formulated as sparse-dense matrix multiplication (SpMM). In natural graphs, edges are usually sparse, represented by a sparse adjacency matrix, while node feature vectors are captured in a dense feature matrix. Aggregating graph data requires multiplying the adjacency matrix with the feature matrix, making this operation a standard SpMM~\cite{huang2020ge}.

SpMM involves numerous irregular computations and random memory accesses, resulting in execution inefficiencies. To mitigate these issues, researchers have proposed various strategies such as optimizing memory access patterns by merging memory accesses and fusing sequential operations~\cite{huang2020ge,rahman2021fusedmm}, utilizing specialized sparse matrix formats and parallel algorithms to address load imbalance~\cite{guo2023bs}, and enhancing sparse kernel performance through preprocessing techniques~\cite{hong2019adaptive,hong2018efficient}. However, these methods primarily focus on standalone SpMM acceleration, and directly applying these approaches to GNNs ignores the coordinated design of sampling and SpMM. ES-SpMM~\cite{lin2021accelerating} leverages the tolerance of pre-trained GNN models to edge loss during inference, exploring SpMM optimization through edge sampling. Nevertheless, due to suboptimal sampling strategies, the method still faces challenges in balancing accuracy and speed for GNN inference. Moreover, while reducing SpMM computation time, it gradually makes data loading the primary bottleneck in GNN inference.

To address these issues, we propose an adaptive edge sampling strategy to accelerate SpMM in GNNs as well as to optimize feature data loading through quantization. Our contributions can be summarized as follows:

\begin{itemize}
	\setlength{\itemsep}{0pt}
	\item[$\bullet$] We identified that existing sampling-based SpMM kernel suffers from an imbalance between accuracy and speed, and that data loading dominates GNN inference time as computation time decreases. We conclude from our analysis that the problem arises from the use of extreme sampling strategies and the memory footprint of the graph feature data.
	\item[$\bullet$] By examining the relationship between the number of non-zero elements in each matrix row and the shared memory width, we introduce AES-SpMM, a SpMM kernel with an adaptive edge sampling approach that effectively balances accuracy and speed. On this basis, we reduce the data loading time with negligible accuracy loss by quantization and dequantization of feature data.
	\item[$\bullet$] Experimental results on graphs of varying scales show that AES-SpMM outperforms both the cuSPARSE SpMM kernel and GE-SpMM by up to 25.87$\times$ and 23.01$\times$, respectively, with less than 1\% accuracy loss. Compared to the sampling-based SpMM kernel ES-SpMM, AES-SpMM reduces accuracy loss by 3.4\% on average and achieves a 1.31$\times$ speedup. Compared to AES-SpMM, quantization-based AES-SpMM has a maximum accuracy loss of 0.3\% and feature data loading time overhead is reduced by 50.91\%-70.51\%.
\end{itemize} 

\section{Background and Motivation}
\label{sec2}

\subsection{GNNs and SpMM}
\label{subsec1}

The runtime of GNN is primarily determined by the aggregation phase, which updates node or edge feature vectors by aggregating neighborhood features, and the combination phase, where a multi-layer perceptron (MLP) extracts high-level features from input features~\cite{zhou2021blockgnn}. Typically, the combination phase employs general matrix multiplication (GEMM), while the aggregation phase relies on SpMM or SpMM-like operations~\cite{gao2022sdma}. The aggregation operation on the graph can be seen as a multiplication between the sparse adjacency matrix and the dense feature matrix, which can be further expressed as $F^{\ell} = A \times H^{\ell}$, where $A$ is the sparse adjacency matrix (representing the graph), $H^{\ell}$ is the dense feature matrix of the $\ell$ layer(representing the node feature), and $F^{\ell}$ is the output feature matrix of the $\ell$ layer. Due to frequent random memory accesses and the lack of hardware-optimized architectures and memory systems for SpMM, the aggregation phase emerges as the primary bottleneck in GNN execution~\cite{gao2023algorithm,liu2023rsc}.

\subsection{Compressed sparse row (CSR) format}
\label{subsec2}

\begin{figure}[t]
	\centering
	\includegraphics[width=0.8\linewidth]{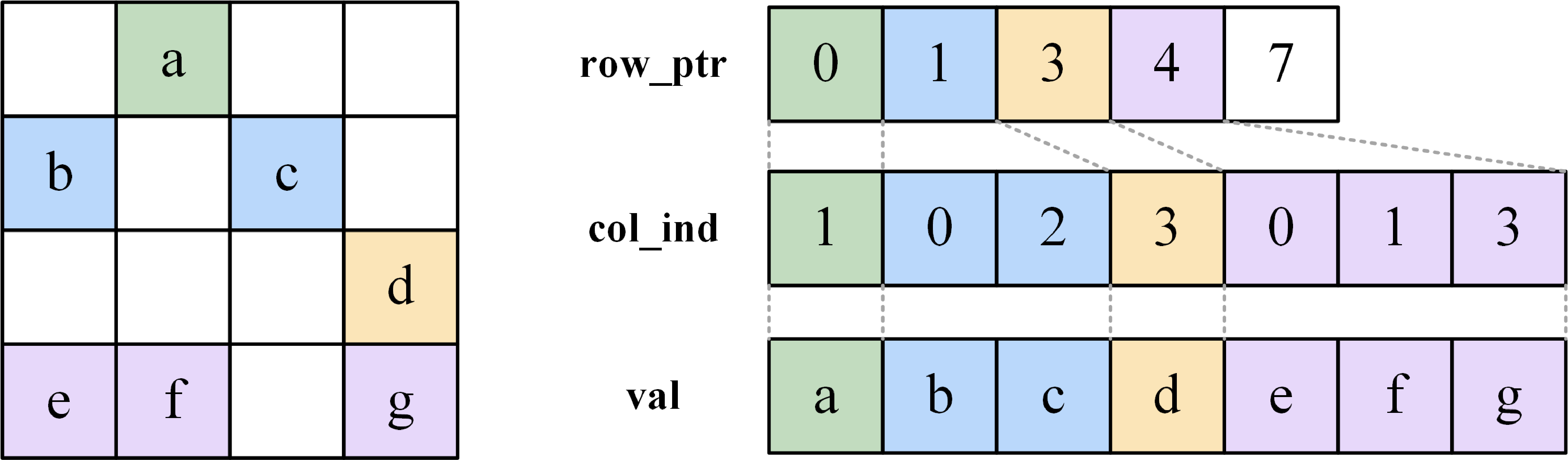}
	\caption{The sparse matrix (left) and its CSR representation (right).}\label{fig1}
\end{figure}

SpMM involves matrix multiplication with sparse matrices, which contain numerous redundant zero elements. Therefore, efficient compression and representation of sparse matrices are crucial. The compressed sparse row (CSR) format is widely used in vendor libraries (such as NVIDIA cuSPARSE~\cite{naumov2010cusparse}) and GNN frameworks (such as DGL~\cite{wang2019deep} and Gunrock~\cite{wang2016gunrock}). As shown in Fig.~\ref{fig1}, the CSR format stores only non-zero elements using three arrays: row\_ptr, col\_ind, and val. The col\_ind and val arrays record the column indices and values of non-zero elements in row-major order, while the row\_ptr array tracks the cumulative count of non-zero elements in each row.

Due to the efficiency and versatility of the CSR format, the AES-SpMM kernel directly adopts this representation. This approach eliminates overhead from additional format conversion, ensuring the operational efficiency of AES-SpMM.

\subsection{GNN feature quantization}

In GNNs, feature data typically constitute the majority of the total data volume. Compressing feature data can effectively reduce data size and improve transmission efficiency. As an efficient data compression technique, quantization has gained increasing popularity~\cite{10.1145/3503221.3508408,9288186,10.5555/3540261.3540777}. Quantization maps Float32 feature values to a limited set of discrete values, representing them using low-bit-width integers such as INT8.

Quantization techniques can be categorized into scalar quantization and vector quantization. Scalar quantization has higher computational efficiency and lower compression and decompression overheads, while vector quantization can achieve higher compression rates but has larger computational overheads. Considering the computational efficiency of the whole inference process, we adopt a lightweight scalar quantization method to reduce the data loading and storage overhead while ensuring accuracy.

Specifically, each feature is quantized and stored using a $b$-bit integer. The scalar quantization process can be calculated as follows:

\begin{equation}
	q = \left\lfloor \frac{x - x_{\min}}{x_{\max} - x_{\min}} \times (2^b - 1) \right\rfloor\label{equ1}
\end{equation}

where $x$ represents the original feature value, $x_{\min}$ and $x_{\max}$ denote the minimum and maximum values of the feature set, respectively, $q$ is the quantized value, and $\left\lfloor \cdot \right\rfloor$ denotes the floor function. After data loading is completed, the quantized value $q$ is dequantized to recover an approximate feature value $\hat{x}$ as follows:

\begin{equation}
	\hat{x} = q \times \frac{x_{\max} - x_{\min}}{2^b - 1} + x_{\min}\label{equ2}
\end{equation}

Since continuous feature values are mapped to a limited set of discrete levels, this process is lossy compression. Therefore, in dequantization, $\hat{x}$ can only approximate the original feature value $x$.

\label{subsec3}

\subsection{Motivation}
\label{subsec4}

While edge sampling has been introduced in GNNs to accelerate SpMM, existing strategies primarily emphasize either accuracy or speed in GNN inference. This unbalanced approach fails to consider that accuracy and speed are equally critical and must be co-optimized for efficient GNN inference. Moreover, as computation time decreases, data loading takes up most of the inference time.

\begin{figure}[t]
	\centering
	\includegraphics[width=\linewidth]{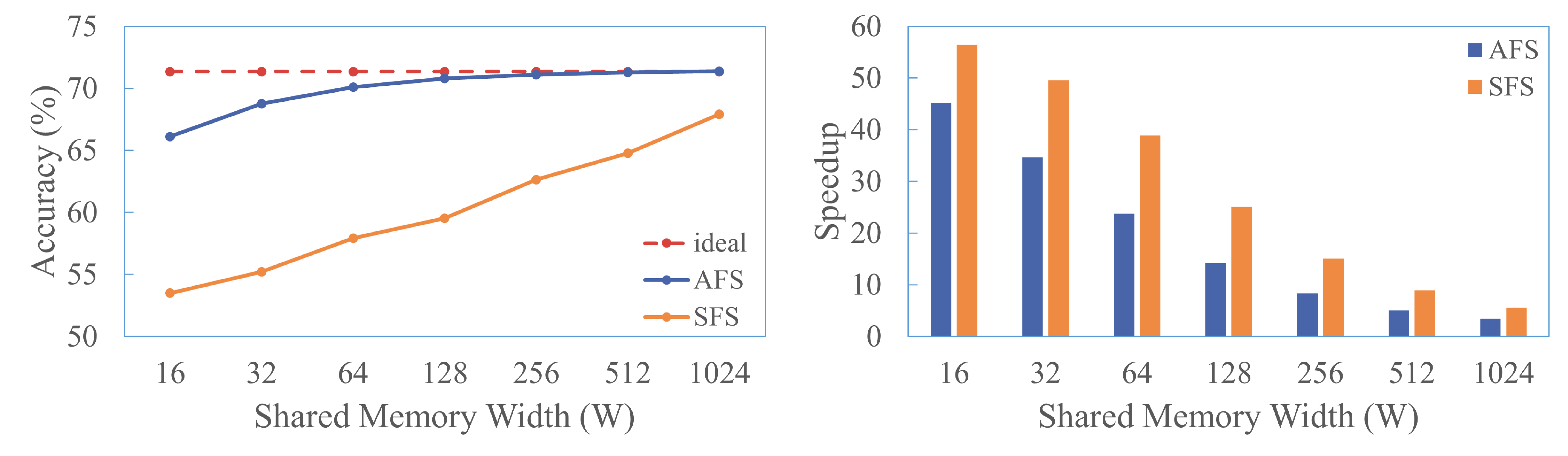}
	\caption{Accuracy (left) and SpMM kernel speedup (right) analysis of the GCN model inference task on the ogbn-proteins dataset using the two strategies of ES-SpMM. AFS represents the accuracy-first strategy and SFS represents the speed-first strategy.}\label{fig2}
\end{figure}

To demonstrate the imbalance between accuracy and speed in GNN inference, we evaluate the performance of GCN using two ES-SpMM strategies on the ogbn-proteins graph dataset and compared it to cuSPARSE SpMM. These two strategies are referred to as the accuracy-first strategy (AFS) and speed-first strategy (SFS). As shown in Fig.~\ref{fig2}, ES-SpMM fits the edge-sampled graph into shared memory, where different sampling levels result in varying accuracy loss and speedup. Specifically, increasing the shared memory width allows for capturing more edge information, but incurs higher sampling overhead. Therefore, as shared memory width grows, GNN inference accuracy increases while speedup decreases. Although AFS achieves higher accuracy than SFS, its speed is suboptimal. Conversely, SFS yields higher speedup but suffers from significant accuracy loss. Comparing AFS and SFS reveals that this imbalance arises from extreme edge sampling strategies. AFS, while ensuring a uniform edge distribution, requires time-consuming calculations for each sampled edge's index. In contrast, SFS speeds up the process by simply judging boundaries, but this leads to substantial loss of graph information due to the concentrated edge distribution.

\begin{figure}[t]
	\centering
	\includegraphics[width=0.9\linewidth]{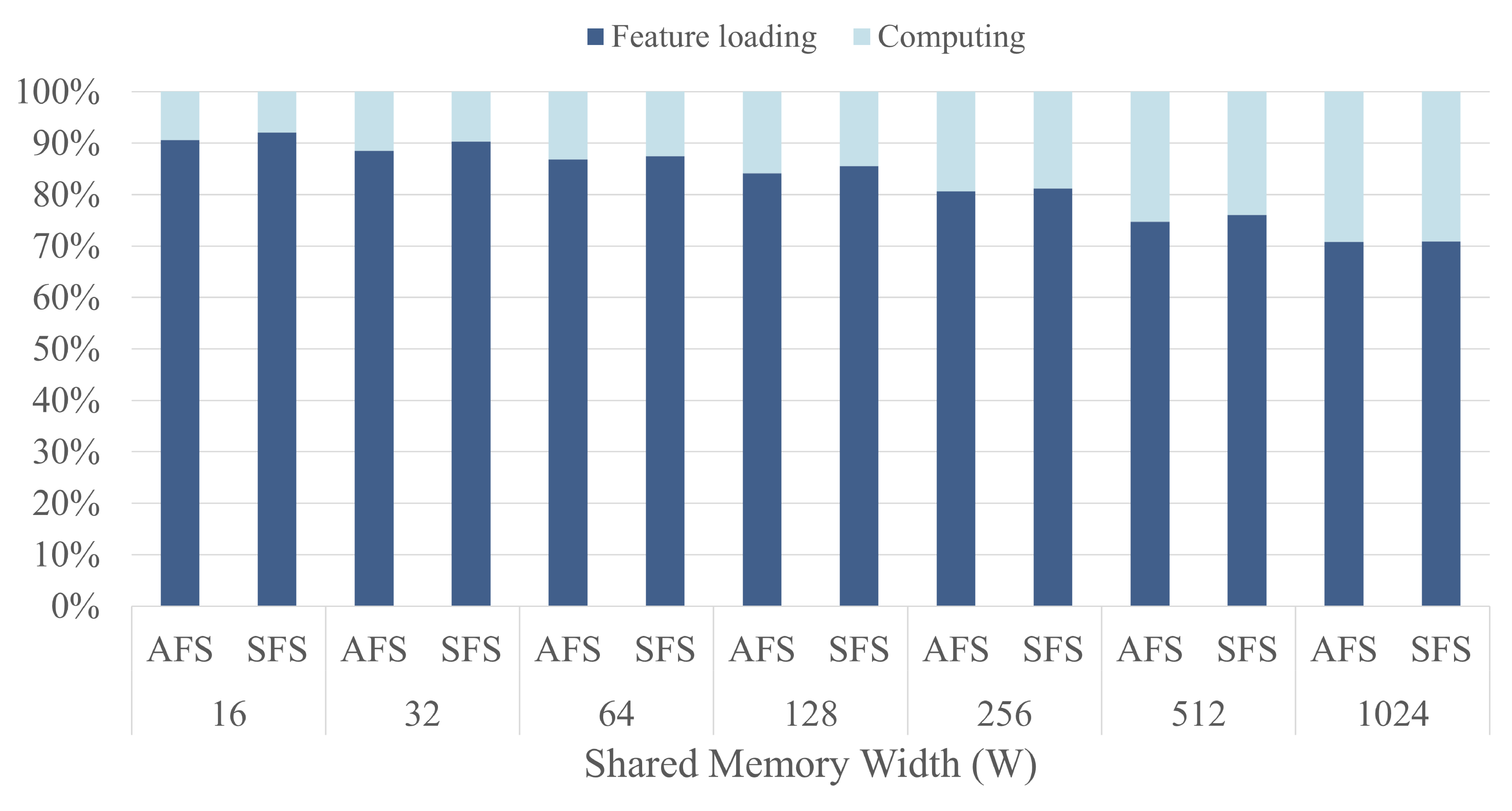}
	\caption{Breakdown of GCN inference time on the reddit dataset using AFS and SFS strategies across different shared memory widths.}\label{fig3}
\end{figure}

To illustrate that feature loading becomes a new bottleneck for GNN inference, we analyze the inference time of GCN model using AFS and SFS on reddit graph dataset. As shown in Fig.~\ref{fig3}, the computing time as a percentage of the total inference time gradually increases as the shared memory width increases. This indicates that the overhead of feature loading dominates when the shared memory width is small, while the growth of computational complexity makes the relative percentage of computing time increase as the shared memory expands. For the same shared memory width, the computing time overhead of AFS is higher than that of SFS, which confirms the previous analysis. Although the computing time percentage rises with the increase of shared memory width, the experimental results show that feature loading is still the main overhead of GNN inference, with the time percentage ranging from 70.78\% to 92.07\%.

To address the above problems, we propose a SpMM kernel that employs an adaptive edge sampling strategy. This strategy selects the sampling scheme based on the relationship between the number of non-zero elements in each matrix row and the shared memory width, thus achieving a balance between accuracy and speed in GNN inference. In addition, we perform INT8 quantization on graph feature data to evaluate the impact of quantization on GNN inference performance when using the proposed SpMM kernel. The method reduces data loading overhead and improves inference efficiency by compressing the data offline before inference and decompressing it efficiently on the GPU end after the feature data is transferred to the GPU.

\section{AES-SpMM Design}
\label{sec3}

\subsection{Feature quantization and dequantization}
\label{subsec5}

The top half of Fig.~\ref{fig4} shows the process of feature quantization. First the quantization process is done offline before GNN inference, and the original graph feature data is quantized to INT8 integers, thus reducing the memory footprint. This process maps each feature value to a discrete quantized value via Equation~\ref{equ1}, compressing the continuous feature values into a smaller representation.

The quantized features are stored in the graph data storage together with the graph structure data and will be loaded during the inference process. Note that in the data loading phase, only the quantized features are loaded instead of the original high-precision data. This makes data loading more efficient and reduces bandwidth requirements especially when dealing with large-scale graph data.

On the GPU end, the quantized features will be restored to approximate the original feature values by performing the dequantization process via Equation~\ref{equ2}. The dequantization process transforms the quantized integers according to the pre-saved $x_{\min}$ and $x_{\max}$, ensuring that the computing process minimizes the loss of precision. The dequantized feature data will be used for subsequent computing and ultimately generate the output required for inference.

\subsection{AES-SpMM kernel design}
\label{subsec6}

\begin{figure*}[t]
	\centering
	\includegraphics[width=\linewidth]{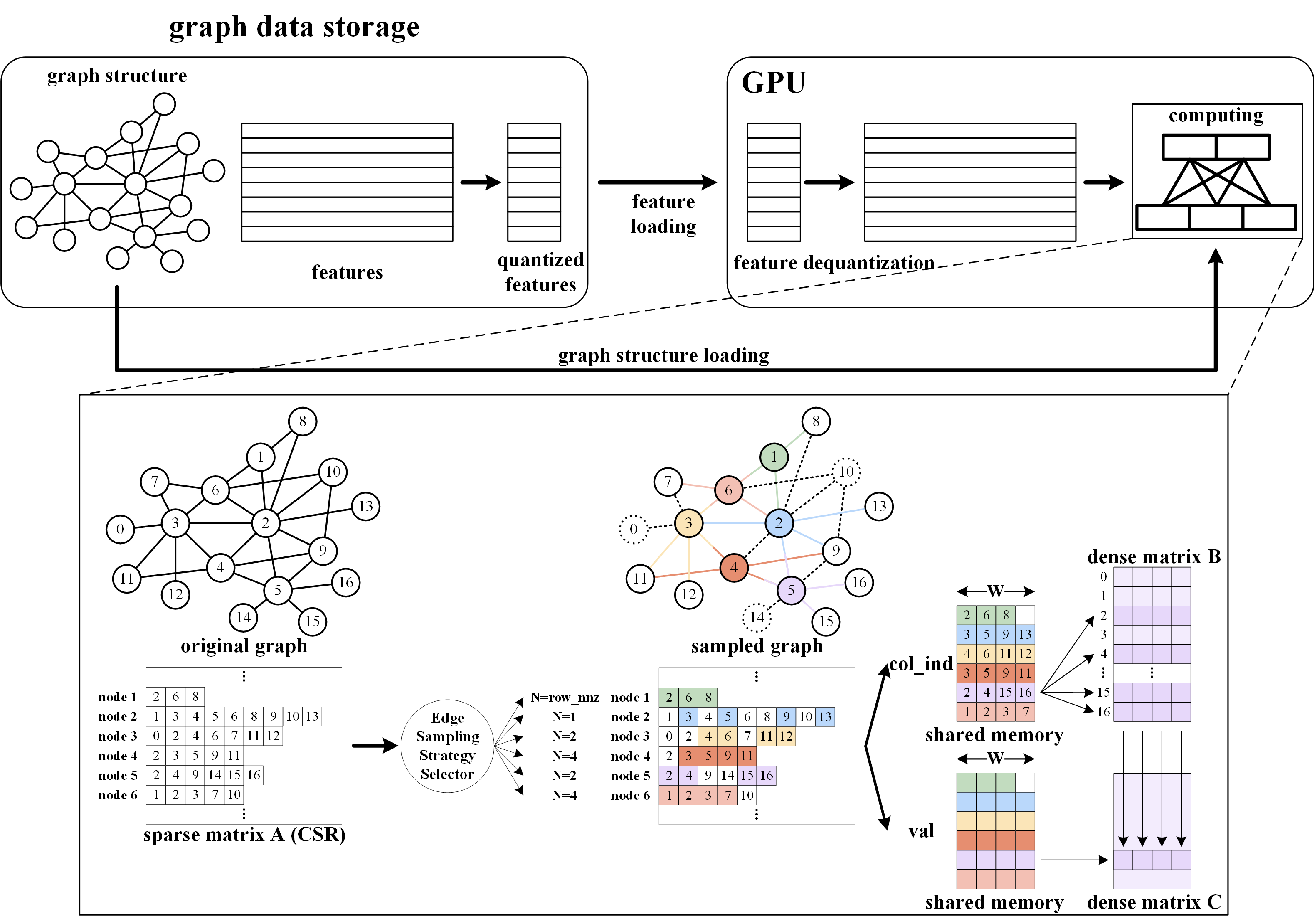}
	\caption{Overview of the feature quantization and dequantization process and the AES-SpMM kernel. The top half shows the feature data processing and the bottom half shows the AES-SpMM kernel, where the sparse matrix A represents the graph structure, the dense matrix B represents the node features, and the dense matrix C represents the output features.}\label{fig4}
\end{figure*}

The workflow shown in the bottom half of Fig.~\ref{fig4} illustrates the operation of the AES-SpMM kernel. Its inputs include the sparse matrix A in CSR format, representing the compressed graph structure, and the dense matrix B, representing the node feature matrix. The objective is to compute the matrix multiplication of A and B, resulting in a dense matrix C that contains the output features. Algorithm~\ref{algorithm1} presents the implementation of AES-SpMM, with a detailed explanation of the kernel workflow provided below.

\begin{algorithm}[t]
	\SetKwData{Left}{left}\SetKwData{This}{this}\SetKwData{Up}{up}
	\SetKwFunction{Union}{Union}\SetKwFunction{FindCompress}{FindCompress}
	\SetKwInOut{Input}{input}\SetKwInOut{Output}{output}
	
	\Input{A.row\_ptr[], A.col\_ind[], A.val[], B[], sh\_width}
	\Output{C[]}
	\BlankLine
	\For{$each\ thread$ \textbf{\emph{in parallel}}}{
		\emph{\_\_shared\_\_ sh\_val[], sh\_col[]}\;
		\emph{sh\_offset, rid, cid, row\_nnz = initialize()}\;
		\emph{result = 0}\;
		\emph{W = min(row\_nnz, sh\_width)}\;
		\emph{N, sample\_cnt = getSampleStrategy(row\_nnz, W)}\;
		\For{$i\leftarrow sh\_offset$ \KwTo $sample\_cnt$}{
			\emph{inc = 0}\;
			\emph{start\_ind = hash(i, row\_nnz, N)}\;
			\For{$j\leftarrow 0$ \KwTo $N$}{
				\emph{sh\_val[i+inc],sh\_col[i+inc]=loadA(start\_ind+j)}\;
				\emph{inc += sample\_cnt}\;
			}
		}
		\emph{\_\_syncthreads()}\;
		\For{$k\leftarrow 0$ \KwTo $W$}{
			\emph{result += sh\_val[k] * B[sh\_col[k], cid]}\;
		}
		\emph{C[rid, cid] = result}\;
	}
	\caption{SpMM with adaptive edge sampling strategy}\label{algorithm1}
\end{algorithm}

First, the shared memory width (W) must be determined based on the graph's complexity and scale. The value of W directly influences both the accuracy and speed of GNN inference. As W represents the number of sampled edges, a larger W captures more comprehensive graph information, but increases edge sampling overhead. For small-scale graphs, even a small W (such as 16) can cover most edges, whereas in large-scale, complex graphs, a small W may lead to significant accuracy loss. After setting W and receiving the sparse matrix A as input, the edge sampling strategy selector in AES-SpMM determines an appropriate sampling strategy for each row. This selection is based on the number of non-zero elements in the row (row\_nnz) and W. If row\_nnz is less than or equal to W, all elements in the row are selected. Otherwise, for row\_nnz larger than W, the selector calculates the ratio (R) of row\_nnz to W and derives the sample count (sample\_cnt) and the number of consecutive elements (N) to sample at a time. Additionally, a hash function computes the starting index (start\_ind) for each sample. The relationship and hash function are detailed in Section~\ref{subsec7}. Finally, AES-SpMM conducts fast adaptive edge sampling using the computed start\_ind and N, storing the selected elements in shared memory. Lines 5 to 14 of Algorithm~\ref{algorithm1} provide the pseudocode for these operations.

Next, AES-SpMM performs the SpMM operation based on the sampling results. Unlike conventional SpMM algorithms, AES-SpMM operates solely on the sampled sparse matrix A. Specifically, it retrieves the values and column indices of A from shared memory, fetches the corresponding values of matrix B from global memory using the column indices, calculates the product of the two values, iterates over this process, and accumulates the results. The final output is matrix C, which contains the computed features. Lines 16 to 19 of Algorithm~\ref{algorithm1} provide the pseudocode for this stage.

Notably, unlike traditional sampling methods and SpMM kernels, AES-SpMM avoids independent and time-consuming sampling. Instead, it employs multiple threads to perform adaptive edge sampling on the rows of matrix A, effectively distributing the task across thousands of threads in parallel. This strategy significantly reduces sampling overhead while maintaining an even distribution of sampled edges. Furthermore, during the SpMM phase, AES-SpMM leverages the tolerance of GNN models to edge loss by storing the sampled A as a sparse matrix in shared memory. This approach minimizes computational cost, enhances data locality, and accelerates SpMM in GNN inference. 

The bottom half of Fig.~\ref{fig4} provides an example to illustrate the AES-SpMM workflow in detail. We set the shared memory width (W) to 4 and input the sparse matrix A in CSR format, focusing on nodes 1 to 6 for simplicity. Since row\_nnz for node 1 is less than W, all its elements are directly selected. For nodes 2 to 6, where row\_nnz is larger than W, the goal is to sample 4 non-zero elements per row for storage in shared memory. The edge sampling strategy selector calculates N based on row\_nnz and W, and determines start\_ind using a hash function. For example, for node 2, N=1, requiring 4/1=4 samples with one element per sample. The computed start\_ind values (1, 3, 6, 8) result in the selection of elements 3, 5, 9 and 13. For node 3, N=2, requiring 4/2=2 samples with two consecutive elements per sample. The start\_ind values (2 and 5) lead to the selection of elements 4, 6, 11, and 12. After completing sampling for all nodes, AES-SpMM transitions to the SpMM stage, where the values from matrix A are multiplied by their corresponding values in matrix B and accumulated iteratively, producing the output matrix C.

\subsection{Adaptive edge sampling strategy}
\label{subsec7}

As outlined above, the adaptive edge sampling strategy in AES-SpMM leverages row\_nnz and W in matrix A to determine sample\_cnt and N for each row based on a predefined relationship. Using a hash function, it calculates start\_ind for each sample, enabling appropriate edge sampling for the current row. The corresponding relationship and hash function will be detailed below.

First, it is critical to identify the factors affecting GNN inference accuracy and speed. Analysis of Fig.~\ref{fig2} and the comparison between AFS and SFS reveal that, with a constant W, fine-grained sampling with a uniform distribution yields higher accuracy but incurs larger computational overheads and slower execution. To balance accuracy and speed, we introduce N, representing the number of consecutive elements, as the sampling granularity. AFS corresponds to fine-grained sampling (N=1), SFS to coarse-grained sampling (N=W), and intermediate values of N offer new sampling possibilities. Table~\ref{tab1} summarizes five sampling strategies, defined by three key parameters: R ( the ratio of row\_nnz to W), N, and sample\_cnt. Specifically, when R is less than or equal to 1 (row\_nnz is less than or equal to W), only one sample is needed, with N=row\_nnz. For R between 1 and 2, over half of the non-zero elements are selected, and only a few are discarded. In this case, four samples suffice, with N=W/4. When R is larger than 54 (row\_nnz is much larger than W, often due to a small W), fine-grained sampling with a uniform distribution ensures accuracy. Here, the strategy involves 32 samples with N=W/32. Other strategies follow a similar pattern, with N constrained to at least 1 and sample\_cnt to at most W in the implementation.

\begin{table}
	\caption{The corresponding relationship in the adaptive edge sampling strategy.}
	\label{tab1}
	\begin{center}
		\begin{tabular}{c c c}
			\hline
			\textbf{\textit{R = row\_nnz / W}} & \textbf{\textit{N}}& \textbf{\textit{sample\_cnt}}\\
			\hline
			$1 \geq R$&row\_nnz&1\\
			$2 \geq R \textgreater 1$&W / 4&4\\
			$36 \geq R \textgreater 2$&W / 8&8\\
			$54 \geq R \textgreater 36$&W / 16&16\\
			$R \textgreater 54$&W / 32&32\\
			\hline
		\end{tabular}
	\end{center}
\end{table}

The design of the hash function is described in Equation~\ref{equ3}. For the selected W elements, the goal is to distribute them evenly across the current row with some randomness to minimize accuracy loss. With N already determined, start\_ind should be chosen pseudo-randomly with large intervals. To achieve both randomness and efficiency, a simple hash function computes start\_ind. As shown in Equation~\ref{equ3}, start\_ind is calculated by multiplying the current index (current\_ind) with a prime number (prime\_num) and taking modulo row\_nnz-N+1. Multiplying current\_ind by a prime reduces the likelihood of repeated start\_ind values, while the modulo operation ensures that start\_ind falls within a valid range. In practice, prime\_num is set to 1429, a large prime that ensures start\_ind spans the full range of row\_nnz.


		\begin{equation}
				start\_ind=(current\_ind\times prime\_num)\bmod(row\_nnz-N+1)\label{equ3}
			\end{equation}

The adaptive edge sampling strategy described above enables AES-SpMM to determine an appropriate sampling granularity for each row, with the sampling task executed in parallel by multiple threads. This strategy is expected to achieve higher accuracy than SFS due to the more distributed sampling of elements, while the speedup over AFS stems from reduced index computation.

\section{Experiment and Evaluation}

\begin{figure*}[h!]
	\centering
	\includegraphics[width=\linewidth]{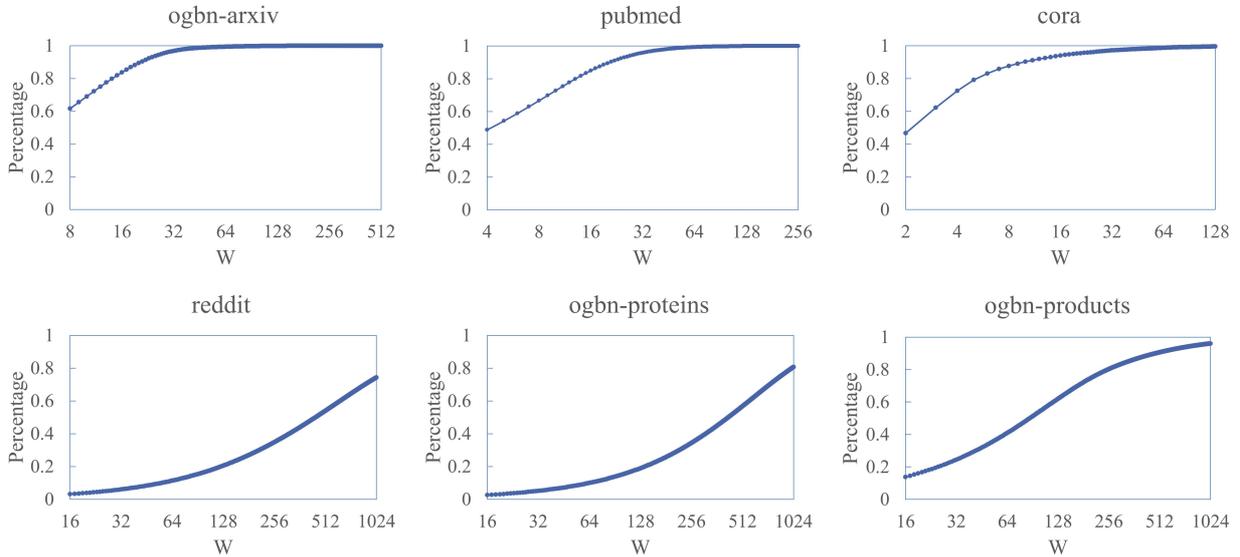}
	\caption{CDF of the sampling rate for AES-SpMM at different W values in different datasets.}\label{fig5}
\end{figure*}

\label{sec4}
\subsection{Experiment setup}
\label{subsec8}


\begin{table}
	\caption{Benchmark graph datasets used for experimental evaluation.}
	\label{tab2}
	\begin{center}
			\begin{tabular}{c c c c c}
				\hline
				\textbf{\textit{Graph}} & \textbf{\textit{Nodes}}& \textbf{\textit{Edges}}& \textbf{\textit{Sparsity (\%)}}& \textbf{\textit{Avg. Degree}}\\
				\hline
				ogbn-arxiv&169,343&1,166,243&0.004067&13.7\\
				pubmed&19,717&88,651&0.022804&4.5\\
				cora&2,708&10,556&0.143947&3.9\\
				reddit&232,965&11,606,919&0.021386&493.0\\
				ogbn-proteins&132,534&39,561,252&0.225224&597.0\\
				ogbn-products&2,449,029&61,859,140&0.001031&50.5\\
				\hline
			\end{tabular}
	\end{center}
\end{table}

\textbf{Platform:} We conducted all experiments on an Intel Core i9-13900KF CPU and an NVIDIA RTX 4090 GPU with 24GB of global memory. This GPU has 128 streaming multiprocessors (SMs). The shared memory size of each SM is 48.0 KB and the maximum number of threads per SM is 1536. The operating system is Ubuntu 20.04. The kernel code is compiled using NVCC 11.8. In our experiments, the execution time includes only the kernel time.

\textbf{Models:} We trained each dataset 10 times using GCN~\cite{kipf2016semi} and GraphSAGE~\cite{hamilton2017inductive} models in the original DGL framework and selected the model with the highest test accuracy. The test accuracy of each selected model is used as the ideal accuracy in the experiments, followed by the inference experiment using the modified DGL framework to call the AES-SpMM kernel.

\textbf{Datasets:} To evaluate AES-SpMM, we used six common benchmark graph datasets, the statistics of which are listed in Table~\ref{tab2}. Ogbn-arxiv represents the citation network for all computer-related papers on the arxiv website. Pubmed is a citation network of medical publications. Cora is a citation network for machine learning papers. Reddit is a collection of social networking data from online forums. Ogbn-proteins use nodes and edges to represent proteins and their biological associations. Ogbn-products represents the co-buying network for Amazon products. Among the six datasets we selected, ogbn-arxiv, pubmed and cora belong to small-scale datasets, while reddit, ogbn-proteins and ogbn-products belong to large-scale datasets.

\textbf{Baselines:} To show the effectiveness of the adaptive edge sampling strategy, we experimentally compare AES-SpMM with the following baselines:

\begin{itemize}
	\setlength{\itemsep}{0pt}
	\item[$\bullet$] cuSPARSE: A vendor's non-open source linear algebra library that is highly optimized for NVIDIA GPUs and supports sparse matrix storage formats such as COO, CSR, and CSC. It is the default implementation in DGL and the main kernel is cusparseSpMM().
	\item[$\bullet$] ES-SpMM: It is an open-source sampling-based SpMM kernel. It implements a cache-first edge sampling architecture and provides two sampling strategies: accuracy-first strategy (AFS) and speed-first strategy (SFS).
	\item[$\bullet$] GE-SpMM: It proposes a novel architecture to accelerate SpMM, which is based on row segmentation design and improves utilization of shared memory in GPU. The efficiency of GNN is improved by introducing the coalesced row caching (CRC) method and the coarse-grained warp merging (CWM) method.
\end{itemize}

\begin{figure*}[h!]
	\centering
	\subfloat[Inference accuracy for GCN.]{
		\includegraphics[width=\textwidth]{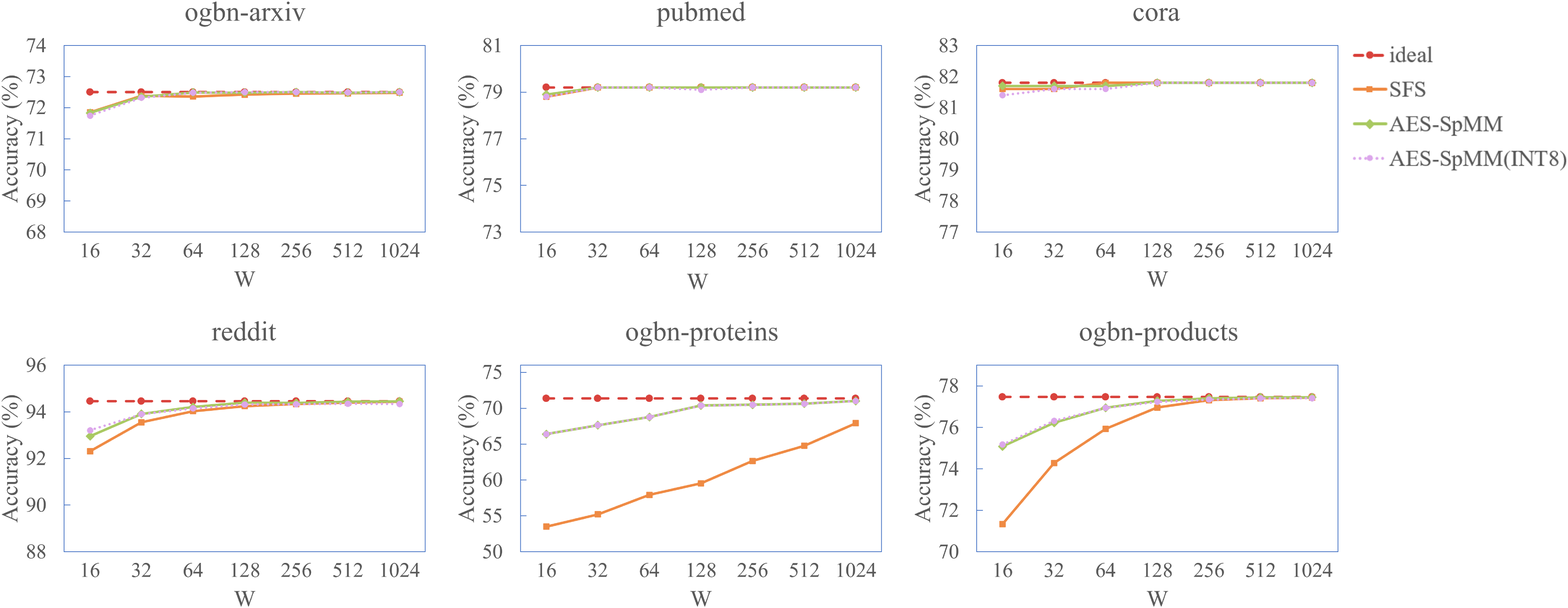}}\\
	\subfloat[Inference accuracy for GraphSAGE.]{
		\includegraphics[width=\textwidth]{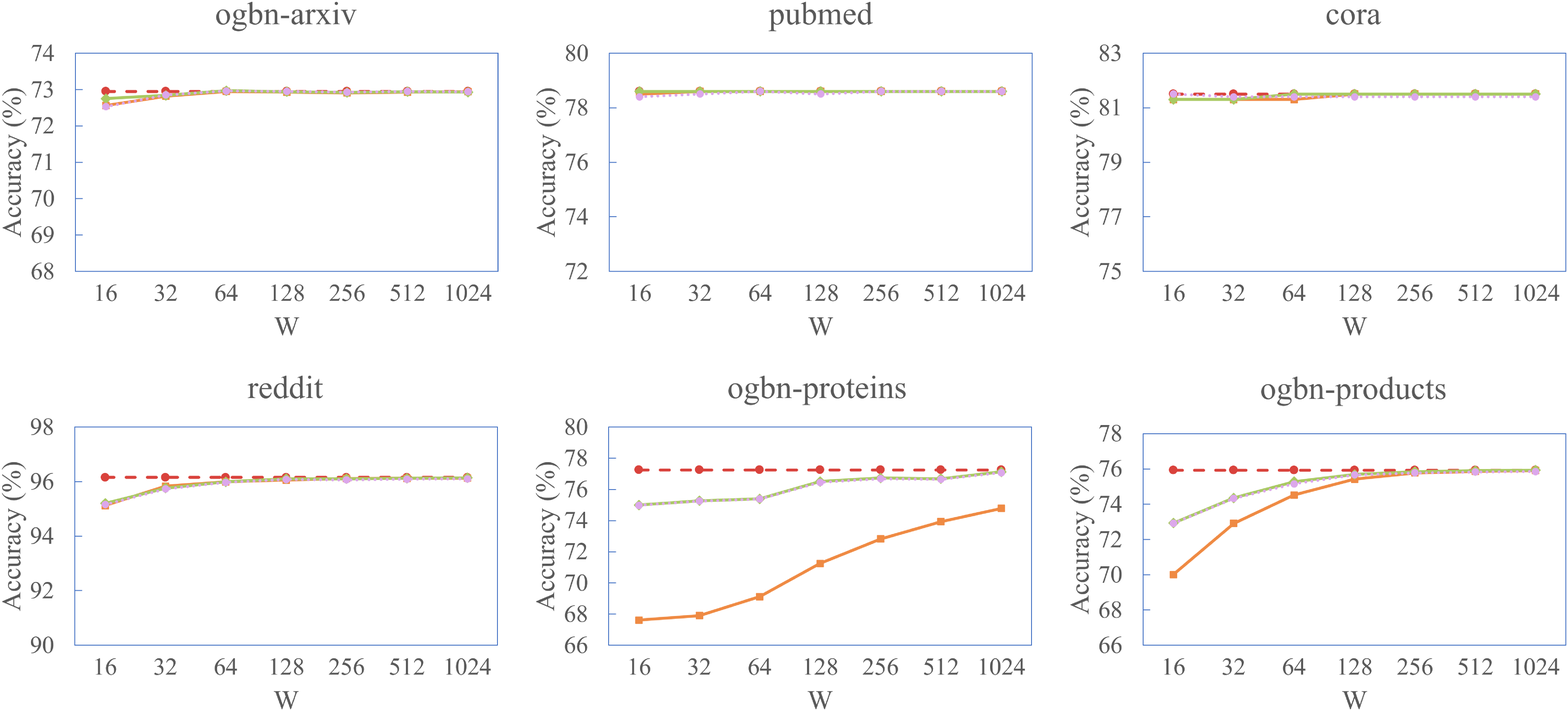}}
	\caption{GCN and GraphSAGE models inference accuracy of AES-SpMM compared to other SpMM kernels on graph datasets of different scales.}
	\label{fig6}
\end{figure*}

\begin{figure*}[h!]
	\centering
	\subfloat[SpMM kernel speedup for GCN.]{
		\includegraphics[width=\textwidth]{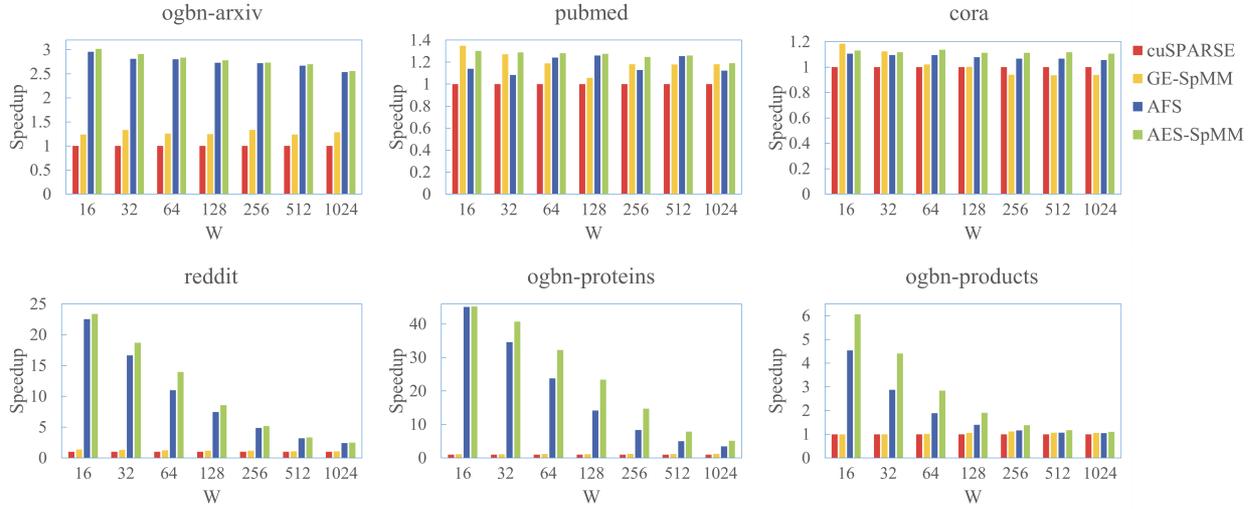}}\\
	\subfloat[SpMM kernel speedup for GraphSAGE.]{
		\includegraphics[width=\textwidth]{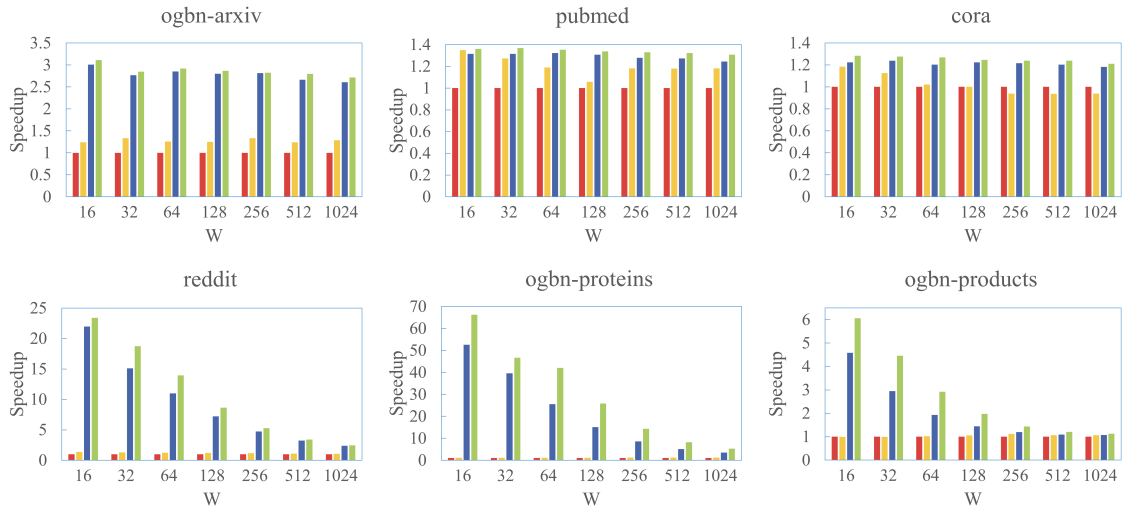}}
	\caption{SpMM kernel speedup of AES-SpMM for GCN and GraphSAGE models compared to other SpMM kernels on graph datasets of different scales.}
	\label{fig7}
\end{figure*}

\subsection{Overall performance analysis}
\label{subsec9}

To evaluate AES-SpMM's performance across graphs of varying scales, we first analyzed its sampling rate under different W values across various datasets. Fig.~\ref{fig5} presents the cumulative distribution function (CDF) of AES-SpMM's sampling rate. As detailed in Section~\ref{subsec6}, for small-scale graphs, even a W value of 16 achieves a sampling rate above 80\%, covering most edges. Conversely, in large-scale graphs, small W values (16 or 32) lead to significant accuracy loss, with sampling rates below 10\%. We trained GCN and GraphSAGE models on six datasets of varying scales using the original DGL framework, measuring inference accuracy and SpMM kernel time for each model. Non-sampling kernels like cuSPARSE SpMM and GE-SpMM incur no accuracy loss during inference. Using the accuracy of cuSPARSE SpMM kernel as the ideal baseline, we modified DGL to replace its kernel with AES-SpMM and conducted inference tests. To analyze AES-SpMM's performance under varying W settings, W values from 16 to 1024 were tested, measuring inference accuracy and SpMM kernel time. For GE-SpMM, the sampling matrices corresponding to different W values were tested, with the best-performing setting selected. Fig.~\ref{fig6} and Fig.~\ref{fig7} compare AES-SpMM's accuracy and speedup against cuSPARSE and open-source baselines.

\subsubsection{Accuracy analysis of AES-SpMM}

Fig.~\ref{fig6}(a) and Fig.~\ref{fig6}(b) compare the inference accuracy of GCN and GraphSAGE models using AES-SpMM and other SpMM kernels. For small-scale graphs, AES-SpMM shows negligible accuracy loss across different W values due to its high sampling rate, even when W=16. In some cases, AES-SpMM achieves slightly higher accuracy than SFS, though the difference is minor, as SFS closely approaches the ideal case. For large-scale graphs, small W values in SFS result in significant accuracy loss, particularly on complex datasets like ogbn-proteins. Even with W=1024, SFS exhibits accuracy losses of 3.4\% and 2.5\%, respectively. In contrast, AES-SpMM significantly improves accuracy, with losses under 1\% for W=128. This improvement stems from more fine-grained edge sampling, which ensures a more uniform distribution of selected elements and allows AES-SpMM to capture more comprehensive information.

\subsubsection{Speedup analysis of AES-SpMM}

Fig.~\ref{fig7}(a) and Fig.~\ref{fig7}(b) present the speedup of GCN and GraphSAGE models using various SpMM kernels, normalized to cuSPARSE. GE-SpMM achieves an average speedup of 1.17$\times$ for small-scale graphs and up to 1.35$\times$ for large-scale graphs. AES-SpMM achieves average speedups of 1.72$\times$ and 1.82$\times$ for GCN and GraphSAGE on small-scale datasets, showing stability due to similar sampling rates across different W values. On small-scale graphs, AES-SpMM provides slightly better speed compared to AFS. For large-scale graphs, as W increases, sampling overhead grows, leading to a general decrease in speedup. However, AES-SpMM demonstrates a clear advantage over AFS. On the ogbn-proteins dataset, GraphSAGE with AES-SpMM achieves a speedup of 25.87$\times$ with less than 1\% accuracy loss, compared to 15.06$\times$ for AFS. This improvement is due to AES-SpMM's coarser-grained sampling strategy, which reduces index computation while effectively controlling accuracy loss, ensuring greater efficiency.

\begin{table}[h]
	\centering
	\scriptsize
	\setlength{\tabcolsep}{10pt}
	\caption{Comparison of feature loading time ratio (\%) among AFS, SFS, and quantization-based AES-SpMM across different models, datasets, and shared memory widths.}\label{tab3}
	\begin{tabular}{c|c|c|c|c|c|c|c|c|c}
		\hline
		\multirow{2}{*}{\centering \textbf{Models}} &\multirow{2}{*}{\centering \textbf{Datasets}} &\multirow{2}{*}{\centering \textbf{Scheme}} & \multicolumn{7}{c}{\centering \textbf{\rule{0pt}{2.4ex}W}} \\ \cline{4-10}
		&&&\rule{0pt}{2.4ex}16&32&64&128&256&512&1024 \\ \hline
		
		&&\rule{0pt}{2.4ex}AFS&55.39 &54.00 &53.88 &54.02 &53.37 &53.39 &52.88 \\ 
		&ogbn-arxiv&SFS&55.46 &54.19 &53.15 &54.37 &53.15 &54.13 &53.57 \\ 
		&&AES-SpMM(INT8)&\textbf{33.64} &\textbf{32.24} &\textbf{32.30} &\textbf{32.84} &\textbf{31.69} &\textbf{31.67} &\textbf{31.80} \\ \cline{2-10}
		
		&&\rule{0pt}{2.4ex}AFS&76.86 &77.19 &76.19 &76.53 &76.41 &76.50 &77.07 \\
		&pubmed&SFS&78.01 &76.06 &75.73 &77.02 &77.66 &76.28 &75.95 \\ 
		&&AES-SpMM(INT8)&\textbf{55.28} &\textbf{54.73} &\textbf{54.11} &\textbf{55.67} &\textbf{55.56} &\textbf{55.72} &\textbf{54.77} \\ \cline{2-10}
		
		&&\rule{0pt}{2.4ex}AFS&61.00 &63.64 &64.71 &62.83 &63.83 &62.05 &63.59 \\
		&cora&SFS&63.40 &62.50 &63.87 &63.96 &63.54 &63.16 &63.18 \\ 
		\multirow{2}{*}{\centering \textbf{GCN}}&&AES-SpMM(INT8)&\textbf{36.36} &\textbf{34.48} &\textbf{38.39} &\textbf{36.28} &\textbf{36.61} &\textbf{36.94} &\textbf{36.13} \\ \cline{2-10}
		
		&&\rule{0pt}{2.4ex}AFS&90.54 &88.50 &86.83 &84.10 &80.68 &74.68 &70.78 \\ 
		&reddit&SFS&92.07 &90.31 &87.42 &85.49 &81.20 &75.98 &70.89 \\ 
		&&AES-SpMM(INT8)&\textbf{76.25} &\textbf{72.73} &\textbf{69.34} &\textbf{64.47} &\textbf{58.37} &\textbf{50.59} &\textbf{44.73} \\ \cline{2-10}
		
		&&\rule{0pt}{2.4ex}AFS&6.49 &6.73 &5.72 &4.71 &3.33 &2.57 &2.06 \\ 
		&ogbn-proteins&SFS&8.37 &7.38 &7.44 &5.99 &5.04 &4.55 &2.61 \\ 
		&&AES-SpMM(INT8)&\textbf{3.00} &\textbf{2.75} &\textbf{2.71} &\textbf{2.49} &\textbf{2.00} &\textbf{1.40} &\textbf{1.08} \\ \cline{2-10}
		
		&&\rule{0pt}{2.4ex}AFS&58.47 &52.66 &46.32 &41.61 &38.15 &36.37 &35.81 \\ 
		&ogbn-products&SFS&64.90 &60.87 &54.19 &46.31 &40.21 &37.62 &36.55 \\ 
		&&AES-SpMM(INT8)&\textbf{34.91} &\textbf{31.33} &\textbf{27.43} &\textbf{22.77} &\textbf{18.90} &\textbf{16.91} &\textbf{16.28} \\   \hline
		
		&&\rule{0pt}{2.4ex}AFS&42.08 &41.87 &41.81 &40.93 &40.87 &40.70 &41.04 \\ 
		&ogbn-arxiv&SFS&41.97 &41.80 &41.31 &41.12 &41.27 &42.06 &41.24 \\ 
		&&AES-SpMM(INT8)&\textbf{23.06} &\textbf{22.91} &\textbf{22.45} &\textbf{22.68} &\textbf{22.78} &\textbf{22.37} &\textbf{21.81} \\ \cline{2-10}
		
		&&\rule{0pt}{2.4ex}AFS&76.99 &75.40 &75.95 &74.87 &76.57 &77.13 &76.41 \\  
		&pubmed&SFS&83.93 &84.18 &85.34 &83.68 &83.93 &83.82 &81.92 \\ 
		&&AES-SpMM(INT8)&\textbf{60.27} &\textbf{57.35} &\textbf{57.55} &\textbf{57.35} &\textbf{57.94} &\textbf{58.49} &\textbf{57.08} \\ \cline{2-10}
		
		&\rule{0pt}{2.4ex}&AFS&69.66 &69.32 &69.89 &69.10 &69.71 &70.06 &68.36 \\ 
		&cora&SFS&70.29 &71.11 &69.94 &71.19 &70.33 &69.83 &68.75 \\ 
		\multirow{2}{*}{\centering \textbf{GraphSAGE}}&&AES-SpMM(INT8)&\textbf{45.65} &\textbf{45.74} &\textbf{45.74} &\textbf{43.96} &\textbf{43.62} &\textbf{47.37} &\textbf{42.39} \\ \cline{2-10}
		
		&\rule{0pt}{2.4ex}&AFS&88.38 &87.14 &85.27 &82.66 &79.60 &74.77 &69.96 \\ 
		&reddit&SFS&89.57 &88.84 &86.86 &83.85 &79.91 &74.52 &70.18 \\ 
		&&AES-SpMM(INT8)&\textbf{73.62} &\textbf{72.31} &\textbf{70.02} &\textbf{65.95} &\textbf{59.74} &\textbf{52.83} &\textbf{47.02} \\ \cline{2-10}
		
		&\rule{0pt}{2.4ex}&AFS&3.89 &3.96 &3.42 &2.93 &2.20 &1.64 &1.27 \\ 
		&ogbn-proteins&SFS&4.18 &4.14 &3.71 &3.63 &3.10 &2.31 &1.78 \\ 
		&&AES-SpMM(INT8)&\textbf{1.93} &\textbf{1.65} &\textbf{1.77} &\textbf{1.54} &\textbf{1.27} &\textbf{1.07} &\textbf{0.76} \\ \cline{2-10}
		
		&\rule{0pt}{2.4ex}&AFS&61.93 &56.57 &49.72 &44.34 &40.48 &39.24 &38.75 \\ 
		&ogbn-products&SFS&69.03 &65.33 &58.41 &49.89 &43.05 &40.72 &39.33 \\ 
		&&AES-SpMM(INT8)&\textbf{38.66} &\textbf{35.40} &\textbf{30.71} &\textbf{24.72} &\textbf{20.80} &\textbf{18.60} &\textbf{18.00} \\   \hline
	\end{tabular}
\end{table}

\subsubsection{Benefits of quantization}

We quantize the feature data to INT8 for storage and loading, and dequantize it to Float32 for computing during the inference process. Fig.~\ref{fig6} illustrates the impact of feature quantization on GNN inference accuracy. Compared to AES-SpMM, quantization-based AES-SpMM brings negligible accuracy loss and even accuracy gain in some cases. On graph datasets of different scales, the average accuracy loss is only 0.02\% for the GCN model, 0.04\% for the GraphSAGE model, and the maximum accuracy loss is only 0.3\% in all test cases.

The main goal of feature quantization is to reduce the feature loading overhead. Table~\ref{tab3} demonstrates the advantages of quantization-based AES-SpMM over AFS and SFS in terms of feature loading time percentage. Since the quantization operation is done in the offline phase before inference, only the additional dequantization overhead needs to be considered during inference. However, since the dequantization operation is executed in parallel on the GPU end, its time overhead is extremely low. Experimental results show that the time overhead of the dequantization process is usually around 2ms, and the quantization-based approach reduces the feature data loading time by 50.91\% to 70.51\% in graph datasets of different scales, which effectively improves the data loading performance.

\vspace{1em}
Overall, AES-SpMM demonstrates better performance across datasets of varying scales. Compared to the non-sampling cuSPARSE SpMM kernel and GE-SpMM, it achieves performance improvements of 25.87$\times$ and 23.01$\times$, respectively, with less than 1\% accuracy loss. For large-scale graphs, AES-SpMM improves accuracy by an average of 3.4\% over SFS and achieves a 1.31$\times$ speedup compared to AFS. By employing an adaptive edge sampling strategy, it optimizes sampling granularity allocation, effectively balancing GNN inference accuracy and speed. Its advantages become even more pronounced with high graph complexity and large-scale. In addition, AES-SpMM introduces a feature quantization mechanism to further optimize the data loading efficiency. Experimental results show that the quantization-based AES-SpMM has a maximum accuracy loss of only 0.3\% in all test cases. Meanwhile, its feature loading time overhead is reduced by 50.91\% to 70.51\%, which further improves the overall inference performance.

\section{Related Work}
\label{sec5}
Issues such as load imbalance and data locality are considered as key performance challenges for SpMM in GNNs. To address these issues, current optimization works can be categorized into software strategy optimization and custom hardware optimization.

\textbf{Software strategy optimization.} SpMM is an important operation in GNN training or inference and dominates the execution time. Several works have been proposed to optimize the execution of SpMM kernel by employing software strategies. TurboMGNN~\cite{10103627} proposes a fine-grained SpMM-based kernel fusion method and implements the fusion of kernels in forward and backward propagation, eliminating redundant accesses to graph data and improving GPU resource utilization. Fan et al.~\cite{10177444} design a unified hybrid parallel strategy for SpMM and SDDMM to achieve load balancing. They propose DTP to adaptively adjust task allocation and HVMA to support aligned, coalesced and vectorized memory access. GE-SpMM~\cite{huang2020ge} efficiently utilizes bandwidth and improves instruction-level parallelism by CRC and CWM. RSC~\cite{liu2023rsc} replaces expensive sparse operations with faster approximate versions and proposes a caching mechanism that reduces the cost of sampling sparse matrices by reusing previous results. DA-SpMM~\cite{10.1145/3489517.3530508} constructs a complete optimization space for SpMM and dynamically selects a better performing design for SpMM by considering the input dynamics. Furthermore, PckGNN~\cite{10579237} and Bs-SpMM~\cite{guo2023bs} design new data storage formats to implement their proposed methods and ensure efficient execution. Our previous work~\cite{song2024efficient} proposes an adaptive edge sampling strategy to address the imbalance between accuracy and speed that exists in sampling-based SpMM kernel. Building on our previous work, this study combines the feature quantization method to further optimize the performance of GNN inference. The quantization effectively reduces the time overhead of data loading while maintaining high inference accuracy by quantizing the feature data offline into low-bit-width integers and dequantizing them at inference time. This work extends the applicability of the proposed method by incorporating new model and new datasets, providing a more comprehensive evaluation and further optimizing the feature loading efficiency in combination with the introduction of quantization, which benefits future research in coordinating design sampling and SpMM in GNN. Compared to these methods, AES-SpMM uniquely integrates sampling-computation coordination design and employs an adaptive edge sampling strategy to balance accuracy and speed.

\textbf{Custom hardware optimization.} A lot of work has been done on SpMM custom acelerators over the past few years to achieve performance and energy efficiency gains. For example, Gao et al.~\cite{gao2023algorithm} propose a dataflow-efficient SpMM algorithm in the algorithm part, and customize accelerator following adaptive dataflow in the hardware part, this algorithm/hardware co-optimization design accelerates SpMM together. SkeletonGCN~\cite{10035119} proposes a unified hardware architecture to process SpMM and dense matrix multiplication (MM) and simplifies computation by compressing scheduling, quantization, and reusing intermediate results. Graph-OPU~\cite{10296283} proposes a GNN-specific ISA and microarchitecture optimized for computation and data communication to improve the overall efficiency and designs a unified matrix multiplication for SpMM and GEMM based on the optimized PCOO format. In contrast, AES-SpMM provides a solution based on general-purpose hardware, delivering scalable performance improvements across different datasets and models without the need for any custom hardware.

\section{Conclusion}
\label{sec6}
In this paper, we present AES-SpMM, a method to enhance SpMM performance in GNNs by balancing accuracy and speed. Our approach addresses the imbalance in existing sampling-based SpMM kernels caused by extreme sampling strategies. We propose an adaptive edge sampling strategy that optimally allocates parallel tasks to GPU resources, incorporating rational sampling granularity and coordinated sampling-computation design. In addition, to further alleviate the data loading bottleneck during GNN inference, we introduce a feature quantization mechanism. By quantizing the features into low-bit-width integers in the offline phase and efficiently performing the dequantization operation on the GPU, the time overhead of feature loading is significantly reduced while maintaining a high inference accuracy. Experimental results show that AES-SpMM effectively balances accuracy and speed, further reduces the feature loading cost when combined with quantization, and comprehensively improves the overall performance of SpMM in GNN inference tasks.

\section{Acknowledgments} 
\label{sec7}
This work is supported financially by Sichuan Natural Science Foundation for Distinguished Young Scholar (ID: 2023NSFSC1966), National Natural Science Foundation of China (ID: 61672438).

\bibliographystyle{elsarticle-num}
\bibliography{bib.bib}

\end{document}